\documentstyle[prl,twocolumn,aps]{revtex}
\input{epsf}
\begin{document}
\draft
\twocolumn[\hsize\textwidth\columnwidth\hsize\csname @twocolumnfalse\endcsname
\title{Exponential Gain in Quantum Computing of Quantum Chaos and Localization}

\author{B. Georgeot and D. L. Shepelyansky}

\address {Laboratoire de Physique Quantique, UMR 5626 du CNRS, 
Universit\'e Paul Sabatier, F-31062 Toulouse Cedex 4, France}

%\date{\today}
\date{October 2, 2000}

\maketitle

\begin{abstract}

We present a quantum algorithm which simulates the quantum kicked rotator model
exponentially faster than classical algorithms.  This shows that important
physical problems of quantum
chaos, localization and Anderson transition can be modelled efficiently on a
quantum computer.  We also show that a similar 
algorithm simulates efficiently classical chaos in certain area-preserving maps.

\end{abstract}
\vskip1pc]

%\begin{multicols}{2}
\narrowtext

The massive parallelism of quantum evolution allows to manipulate
simultaneously exponentially many states through entanglement (see reviews
\cite{josza,steane}).  That opens new
horizons for computations based on quantum mechanics, as was stressed by
Feynman \cite{feynman}.  Nevertheless it is not so obvious if this
parallelism can be used to speed up exponentially any given computational
algorithm.  There is certainly no systematic way to do this, and that is why so
much interest has been generated by the Shor quantum algorithm \cite{shor} which
factorizes large numbers exponentially faster than any known classical
algorithm.  At present very few other quantum algorithms have been found 
in which substantial
computational gain is achieved compared to classical computing.  Among them is
the Grover algorithm \cite{grover} which significantly accelerates the problem
of searching an unsorted database, although the gain is not exponential.  In
addition quantum computers can be used as analog machines to simulate some
many-body quantum systems which are hardly accessible in usual computer
simulations \cite{lloyd}.  In this way some systems such as spin lattices can 
be modelled very naturally for example by cold atoms in optical 
lattices \cite{molmer}.  However this kind of simulation is restricted
to systems whose physical elements are related or similar 
to the qubits (spin halves)
 of a quantum computer.  At present apart from these natural examples 
 there are no developed quantum algorithms
 which would allow to reach exponential gain in the computation of the quantum
dynamics of physical systems.

It is therefore desirable to find a quantum algorithm corresponding to a
physical model with rich and complex quantum dynamics.  During the last
decades, it has been understood that generally the dynamics of classical
nonlinear systems is chaotic \cite{lichtenberg}.  The corresponding quantum 
dynamics, called
quantum chaos, demonstrates a rich and complex behaviour even for systems with
only few degrees of freedom and rather simple Hamiltonians \cite{houches}.
One of the cornerstone models in the study of quantum chaos is the kicked
rotator.  In the classical limit, this model reduces to an area-preserving map
called the Chirikov standard map \cite{chirikov} which has applications in different fields
of physics, such as particle confinement in magnetic traps, beam dynamics in
accelerators, comet trajectories and many others \cite{lichtenberg}.  The map
depends on only one parameter, and depending on its value the system can be in
the near-integrable bounded r\'egime, with Kolmogorov-Arnold-Moser (KAM) curves, or
in the fully chaotic r\'egime with diffusive growth of momentum which
statistically can be described by the Fokker-Planck equation.  In between these
two r\'egimes the phase space of the system has a complex hierarchical
structure with integrable islands surrounded by a chaotic sea at smaller and
smaller scales.  The quantum dynamics corresponding to these different
r\'egimes have been intensively studied by different groups in the field of
quantum chaos \cite{kicked}.  Many phenomena of general importance are present
in this model, including quantum ergodicity, spectral statistics as in 
random matrix theory, quantum KAM r\'egime, and many others.  However the most
unexpected quantum effect is the phenomenon of dynamical localization, in which
quantum interference suppresses chaotic diffusion in momentum, leading to
exponentially localized eigenstates.  This effect has close analogy with
Anderson localization of electrons in disordered materials \cite{fishman}, and
therefore this model enables to study also the properties of Anderson
localization, a solid-state problem still under intense investigation nowadays.
The quantum kicked rotator describes also the properties of microwave
ionization of Rydberg atoms \cite{IEEE}.  It
has been realized experimentally with cold atoms, 
and the effects of dynamical localization, external noise and decoherence
have been studied experimentally \cite{raizen}.

In this paper we present a quantum algorithm which computes the
evolution of the quantum kicked rotator exponentially faster than any classical
computation.  It simulates the kicked rotator with $N$ levels
in $O((\log_{2} N)^3)$ operations
instead of $O(N\log_{2} N)$ for the classical algorithm.

The classical dynamics of the system is given by the Chirikov standard map
\begin{equation}
\label{map}
\bar{n} = n + k \sin{  \theta }; \;\;\; 
\bar{\theta} = \theta + T \bar{n}
\end{equation}
where $(n,\theta)$ is the pair of conjugated momentum (action) and angle 
variables, and the bars denote the resulting variables after one iteration of the
map.  In this way the dynamics develops on a cylinder (periodicity in $\theta$)
which can be also closed to form a torus of length $N=2 \pi L/T$ where
$L$ is an integer.
The classical dynamics depends only on one single chaos parameter $K=kT$,
so that the motion is globally chaotic for $K >  0.9716...$  
For $K \gg 1$ the orbits spread diffusively in $n$ with diffusion rate $D=n^2/t
\approx k^2/2 $ where $t$ is measured in number of iterations (kicks)
\cite{chirikov,lichtenberg}.  In the chaos regime, the dynamics is
characterized by positive Kolmogorov-Sinai entropy $h \approx \ln (K/2)>0$, 
 due to which trajectories diverge exponentially and
roundoff errors grow exponentially with $t$ 
\cite{dls83}.  

The quantum evolution during one period is described by an unitary operator
acting on the wave function $\psi$
\begin{eqnarray} 
\label{qmap}
\bar{\psi} = \hat{U} \psi =  e^{-ik\cos{\hat{\theta}}}
 e^{-iT\hat{n}^2/2} \psi,
\end{eqnarray}
where $\hat{n}=-i \partial / \partial \theta $, $\hbar=1$.  In this way the 
classical limit corresponds to $k \rightarrow \infty$, $T \rightarrow 0$ while
keeping $K=kT=const$ \cite{kicked}.  The quantum interference leads
to exponential localization of the eigenstates $\chi_m(n)$ 
of the operator $\hat{U}$ in the
momentum space $n$ with envelope $\chi_m(n) \sim \exp(-|n-m|/l)/\sqrt{l}$ 
where $m$ marks also
the center of the eigenstate and $l$ is the localization length.  In the 
r\'egime of quantum chaos ($k \gg K \gg 1$) this length is determined by the 
classical diffusion rate $l = D/2 \approx k^2/4$ \cite{kicked}. The evolution
takes place on $N$ levels with periodic boundary conditions.  For $l \gg N$ 
the eigenstates become ergodic and 
the level spacing statistics is described by random matrix theory \cite{kicked}.
Therefore depending on the parameters various regimes of quantum chaos can be
investigated in this single model.

The evolution operator $\hat{U}$ is the product of two unitary
operators $\hat{U}_{k}=\exp(-ik\cos{\hat{\theta}})$ and 
$\hat{U}_{T}=\exp(-iT\hat{n}^2/2)$ which represent respectively
the effects of a kick and free rotation.  These operators are diagonal in
the $\theta$ and $n$ representations respectively.  Due to that the most 
efficient way to simulate the quantum  dynamics of this system on a classical
computer is to perform forward/backward fast Fourier transforms (FFT) to go
from one representation to the other \cite{note0}, 
doing diagonal multiplications by
$U_k$ and $U_T$ between each FFT.  In this way for a system with $N$ levels
the FFT requires $O(N\log_2(N))$ operations and the diagonal multiplications
take $O(N)$ operations, so that evolution on one period is performed in 
$O(N\log_2(N))$ operations \cite{note1}.  Our construction of the quantum 
algorithm keeps the global structure of this classical algorithm, and uses
quantum parallelism to speed up exponentially each algorithmic step.

{\em Step I: Preparation of the input state.} We consider a system 
of $n_q$ qubits; the Hilbert space of dimension $N=2^{n_q}$ is used
to describe $N$ momentum states
(eigenstates of the operator $\hat{U}_{T}$ for $0 \leq n \leq N-1$ 
in binary code) on which evolves the kicked rotator.  
An initial state $\psi(0)= \sum_{n=0}^{N-1} a_n |n>$ 
at time $t=0$ is prepared by rotations
of individual qubits and two-qubit gates from the ground state $|0...0>$. 
For example, a typical 
initial state used in the studies of the kicked rotator dynamics \cite{kicked},
such as $\psi(0) = |N/2>$, requires only one individual rotation.
 We need also
auxiliary registers which will be used later; at the moment they are 
all in the ground state.

{\em Step II: Action of free propagation operator $\hat{U}_{T}$.}  
In the $n$ representation $\hat{U}_{T}$ is diagonal so that 
$\hat{U}_{T} |n>=\exp(-iTn^2/2) |n>$.  The simultaneous multiplication of the
$N$ coefficients can be done in $n_q^2$ gate operations.  Indeed, if
$n=\sum_{j=0}^{n_q-1} \alpha_j 2^j$, then
$n^2= \sum_{j_1,j_2} \alpha_{j_1} \alpha_{j_2} 2^{j_1+j_2}$.  Therefore
$\exp(-iTn^2/2)=\Pi_{j_1,j_2} \exp(-iT\alpha_{j_1} \alpha_{j_2} 2^{j_1+j_2-1})$ 
with $\alpha_{j_{1,2}}=0$ or $1$.  As a result, this step can be realized with 
$n_q^2$ operations of the two-qubit gate applied to each qubit pair 
$(j_1,j_2)$ which keeps the states $|00>,|01>,|10>$ 
unchanged while $|11>$ is transformed to $\exp(-iT 2^{j_1+j_2-1}) |11>$ 
\cite{division}.

{\em Step III: Change from $n$ to $\theta$ representation.}  In analogy
with the classical algorithm, we can use the quantum Fourier transform (QFT)
(described in detail, for example in \cite{josza}). The QFT requires
$O(n_q^2)$ operations with one-qubit rotations and two-qubits gates similar
to the ones described above.  After the QFT, we obtain the wave function
in the $\theta$ representation, $\sum_{i=0}^{N-1} b_i |\theta_i>$, where the
$\theta_i$ are the binary codes of $N=2^{n_q}$ discretized angles.  We note
that the QFT was discussed in \cite{schack} for simulating a
rather specific model, the quantum baker map.

{\em Step IV: Construction of a supplementary register holding the 
cosines of angles}.  This step transforms 
$\sum_{i=0}^{N-1} b_i |\theta_i>|0>$ into 
$\sum_{i=0}^{N-1} b_i |\theta_i>|\cos\theta_i>$.  The second register 
is of course present since step I, but is used only at that step.  After these
operations, it contains the binary codes of the $N$ values of $\cos\theta_i$, 
correlated with $\theta_i$ in the first register.  The number of qubits $p$
in this second register sets the precision of the cosines at $2^{-p}$, 
and should be equal or greater than $n_q$.  This register will be used
in the next step to perform the kick operator $\hat{U}_{k}$.  
To realize this transformation, we need a few auxiliary registers which can
be erased at the end.  First we precompute the $2n_q$ values
$\cos (2\pi/2^j), \sin (2\pi/2^j)$, for $j=1,..,n_q$ with precision $2^{-p}$.
This can be done quantum mechanically or classically in polynomial time
by first computing the case of smallest angle
and then using recursive relations, doubling the angle each time.  
Also, other classical methods converging superexponentially (e. g. Newton's)
can be used.  We decompose $\theta_i$ in binary code  
$\theta_i =\sum_{j=1}^{n_q} \beta_{ij} 2\pi/2^{j}$ and
use the formula $\exp(i\theta_i) = \Pi_{j=1}^{n_q} \exp(i\beta_{ij} 2\pi/2^{j})
=\Pi_{j=1}^{n_q} (\cos(\beta_{ij} 2\pi/2^{j}) +i \sin(\beta_{ij} 2\pi/2^{j}))$,
with $\beta_{ij}=0$ or $1$, to compute $\cos\theta_i$ and $\sin\theta_i$
 in $4n_q$ multiplications.  This can be done in parallel for all 
$N$ values of $\theta_i$ in $O(n_q p^2)$ gate operations. 
We need for that an auxiliary register
with $p$ qubits on which $\sin\theta_i$ is built.  We start therefore
with $\sum_{i=0}^{N-1} b_i |\theta_i>|1>|0>$, then we perform 
the transformation $\sum_{i=0}^{N-1} b_i |\theta_i>|c>|d> \rightarrow
\sum_{i=0}^{N-1} b_i |\theta_i>|\cos (\beta_{ij}2\pi/2^j)c-
\sin (\beta_{ij}2\pi/2^j)d>
|\sin (\beta_{ij}2\pi/2^j)c+\cos (\beta_{ij}2\pi/2^j)d>$ for $j=1,..,n_q$ 
(initially $|c>=|1>, |d>=|0>$).  This transformation needs a controlled multiplier
with the qubit $j$ as control qubit.  The quantum circuits realizing a
controlled multiplier are described in \cite{vedral}, and require
$O(p^2)$ gate operations for each multiplier.  
After $n_q$ transformations we obtain 
$\sum_{i=0}^{N-1} b_i |\theta_i>|\cos \theta_i>|\sin \theta_i>$.
The total number of gate
operations for step IV is therefore $O(n_q p^2)$.  We note again that $p$ 
determines the precision with which the $\cos\theta_i$ are computed.  A
reasonable r\'egime is $n_q \leq p \leq 2n_q$ which gives a total number
of gate operations for this step of $O(n_q^3)$.
Besides the three registers 
already described, some auxiliary registers 
are necessary to perform these operations.  It can be done with $5$ additional
registers of size $p$, but it is probably possible to decrease this number.  
After that
only the registers holding $\theta_i$ and $\cos\theta_i$ will be used, and
all others can be erased.  The auxiliary registers holding intermediate
values of cosines and sines can be reused
after each multiplication by $\exp(i\beta_{ij} 2\pi/2^{j})$ is performed,
since they can be erased by using multiplication by 
$\exp(-i\beta_{ij} 2\pi/2^{j})$ and subtraction.

{\em Step V: Action of kick operator $\hat{U}_k$}.  After the previous steps,
the state of the system is 
$ \psi=\sum_{i=0}^{N-1} b_i |\theta_i>|\cos\theta_i>$.  In the angle 
representation, the action of $\hat{U}_k$ is diagonal so that 
$\hat{U}_k |\theta_i>= \exp(-ik\cos\theta_i) |\theta_i>$.  Each state
$|\theta_i>$ is entangled with $|\cos\theta_i>$ holding the binary code of 
$\cos\theta_i = \sum_{j=1}^{p} \gamma_{ij} 2^{-j}$, with $\gamma_{ij}=0$ or
$1$.  Since $\exp(-ik\cos{\theta_i})= \Pi_{j=1}^{p} \exp(-ik \gamma_{ij}
2^{-j})$, to perform the multiplication, it is therefore 
enough to apply to each qubit of the second register the one-qubit 
gate which takes $|0>$ to $|0>$ and $|1>$ to $\exp(-ik2^{-j})|1>$. Only
$p$ gate operations (with $n_q \leq p \leq 2n_q$) are used for this
transformation. After this $|\cos\theta_i>$ is reversibly erased.
As a result of this step, the state of the system is now
$\sum_{i=0}^{N-1} b_i \exp(-ik\cos\theta_i)|\theta_i>|0>$.

{\em Step VI: Change from $\theta$ to $n$ representation}.  This step
is similar to step III, it performs backwards the QFT on the first register
$|\theta_i>$ and returns the wave
function to the momentum basis in $O(n_q^2)$ operations.
This gives the wave function of the kicked rotator after one iteration of
map (\ref{qmap}) (one kick step).

In this way one kick iteration requires $O(n_q^3)$ gate operations.
Subsequent kicks are realized using
steps II-III-IV-V-VI, since
step I is done only once.   As a result, a quantum
computer can perform the kicked rotator evolution exponentially faster than
a classical computer (which needs $O(n_q 2^{n_q})$ operations per kick).
Several successive measurements after $t$ iterations can give 
the largest probabilities  $|a_i|^2$ 
in the momentum basis from which for example the localization length $l$ can be
extracted with only few measurements.  Other average characteristics can be
 obtained efficiently by performing the QFT followed only by few measurements, 
giving for example the largest harmonics of the probability distribution.

Some modifications in this algorithm are possible.  
Instead of the additional register $|\cos\theta_i>$ one can use a register
holding an uniform polynomial approximation $P(\theta) \approx \cos\theta $. 
This is actually the technique used in classical computation, where Chebychev
polynomials are used.  The construction of the polynomial of degree $p$
is done iteratively, starting from the lowest-degree monom and multiplying
it by $\theta_i$ to obtain the next one.  An additional register
holds the current power of $\theta_i$ between each step.  
The coefficients of the polynomial
should be precomputed in advance.  The transition from one degree to another
requires $O(p^2)$ gate operations for $p$ qubits,
so that the whole process needs $O(p^3)$
operations.  For $p\sim n_q$, this is similar to step IV.  After computing
the register $|P(\theta_i)>$ the next step is unchanged and performs
multiplication by $\exp(-ikP(\theta))$ \cite{note2}.

The generalized models of kicked rotator, where $k\cos\theta$ is replaced
by another function $V(\theta)$, are also of interest.  In general it is not
obvious if the register $|V(\theta_i)>$ in step IV can be computed in
polynomial time, so uniform polynomial approximations can be the only way.
A case of particular interest is $V(\theta)= 2\mbox{arctan}(E-2k\cos\theta)$,
where $E$ and $k$ are parameters.  In this case, the kicked rotator can be
exactly mapped on a solid-state system on a chain with only nearest-neighbour
hopping, as it had been shown in \cite{fishman}.  This computation
can be done via uniform polynomial approximation.   It may also be possible that 
this $V(\theta)$ can be computed directly in parallel from registers
$|\cos\theta>|\sin\theta>$.

The generalized kicked rotator can also model
 the localization properties in higher
dimensions.  For example, if the parameter $E$ varies from kick to kick, for
example $E= -2k\cos(\omega_1 t) - 2k \cos(\omega_2t)$ where $\omega_1/(2\pi)$ 
and $\omega_2/(2\pi)$ are two incommensurate frequencies, then the model can be
mapped on the three-dimensional Anderson model which displays a metal-insulator
transition \cite{casati89}.  For $k < k_c \approx 0.46$ the dynamics is
localized in momentum space, while for $k>k_c$ it becomes diffusive and ergodic
over all the available space of size $N$.  Similar effects were observed
recently in quasiperiodically driven cold atoms \cite{delande}.   However,
this latter model is still operating on a kicked rotator with one degree
of freedom.  One can consider a d-dimensional version of (\ref{qmap}) with
$\hat{U}_k=\exp(-ik\sum_{r=1}^{d} \cos\theta_r)$ and 
$\hat{U}_T=\exp(-iT\sum_{r=1}^{d} n_r(n_r + {\sum_{r<r'}^{d}n_{r'}})/2)$, 
which is directly related to the
$d$-dimensional Anderson localization problem.
The simulation of this model  on a system of size $N=2^{n_q}$ 
for each momentum $n_r$  requires $O(N^d \log_2 N)$ 
operations on a classical computer since the total basis contains $N^d$ levels.
Even for $d=2$ this model is hardly accessible for nowadays computers
\cite{fishman2}. On the contrary, our quantum algorithm can be directly
extended to higher dimensions by increasing the number of registers by a factor
$d$, and the number of operations becomes only $O(dn_q^3)$. 
For $d=2$ by changing $\hat{U}_T$ to 
$\hat{U}_T=\exp(-iT(n^2_1+n^2_2)/2-i g \delta_{n_1,n_2})$ it is possible to  
simulate the problem of two interacting particles 
in a localized phase \cite{dls94}.
We note that usually one needs an exponential number of gates
to simulate a quantum map operator, and a
simulation in polynomial time is not obvious. For example for the 
extensively studied kicked top  \cite{haake} the QFT cannot be used 
to change representations.

It is interesting to note that on the basis of the algorithm constructed above,
one can simulate also the classical map (\ref{map}).  Indeed, the
discretization of the map can be done in a symplectic way \cite{symp} on a
phase space lattice of size $N \times N$:
\begin{equation}
\label{dismap}
\bar{Y} = Y + S_{N} (X); \;\;\; 
\bar{X} = X + \bar{Y} (mod N)
\end{equation}
with  
$S_N(X)$ = $[ N K  \sin (2 \pi X/N)/(2 \pi) ]$,  
$[... ]$ being the integer part  and $X, Y$ integers.  
This map is area-preserving and invertible.
An initial classical phase space density 
can therefore be modelled by a quantum state $\sum_{i_1,i_2} a_{i_1,i_2}
|X_{i_1}>|Y_{i_2}>|0>$.  Then we use step IV to transform it in 
$\sum_{i_1,i_2} a_{i_1,i_2} |X_{i_1}>|Y_{i_2}> |S_N(X_{i_1})>$ in
$O(\log_2(N)^3)$ operations, performed only once at the beginning. After that 
the map is reduced to simple additions in the first two
registers and require only $O(\log_2(N))$ operations per iteration, simulating
exponentially many trajectories in polynomial time.
A simulation on a classical computer requires $O(N^2)$ operations per
iteration if one simulates a density distributed over $O(N^2)$ lattice cells. 
After $t$ iterations  $a_i$ give the density probability distribution in
phase space.  QFT can be used to determine the main harmonics of the density,
 which can be measured by running the algorithm several times.  

In conclusion, the algorithm presented here shows that important systems 
displaying rich and complex properties like classical and quantum chaos,
quantum localization and ergodicity can be simulated exponentially faster
on a quantum computer.  This allows to study the time-evolution of these systems
and the transition between classical and quantum mechanics in the limit of
large quantum numbers (semiclassical limit).  Using a few tens of qubits 
can probe this limit far beyond what is possible on nowadays computers. 
The problem of the precision of such quantum computation is crucial, and
the effects of imperfections should be studied in detail for 
different physical properties of the simulated system 
(some of them can be quite sensitive \cite{song}).

We thank G.Benenti and K.Frahm for stimulating discussions.


\begin{thebibliography}{99}
\bibitem{note0} The use of the FFT automatically puts the system on a torus; to
                be consistent, one should take $T/(4\pi)= M/N$ where $M$ 
                is an integer \cite{kicked}.
\bibitem{note1} Another numerical method is based on the expansion of $U_k$
                in $n$-representation over Bessel functions, avoiding 
                the FFT.  However  this requires about $O(kN)$ operations 
                with Bessel functions that become much less effective 
                than the FFT  in the semiclassical limit $k \gg 1$.
\bibitem{division} These $2n_q$ gates require to precompute
                   $T 2^{j_1+j_2-1}$ $ (\mbox{mod} \; 2\pi)$, which can be done classically
                   through Euclid's algorithm in $O(n_q)$ operations.
\bibitem{note2} It is possible to avoid step IV
                and compute directly $\exp(-ikP(\theta))$ by expanding 
                each power of $\theta$ in the polynomial in binary 
                representation (in a way similar to step II for
                $\exp(-iTn^2/2)$ in momentum representation).  
                This uses only one register of size $n_q$, 
                but requires $O(n_q^p)$ operations, that becomes exponential
                for $p \sim n_q$.
\end{thebibliography}
\end{document}